\documentclass[10pt,notitlepage,oneside,twocolumn]{article}

 \usepackage{cite}

 \usepackage[centertags]{amsmath}
 \usepackage{amsfonts}
 \usepackage{amssymb}
 \usepackage{graphicx}
 \usepackage{lscape}
 \usepackage{amsmath}

 \usepackage{caption2}

\newcommand{\erf}{\mathrm{erf}}
\newcommand{\Tr}{\mathrm{Tr}}

\begin{document}
\title{Schmidt number of pure continuous-variable
bi-partite entangled states and methods of its calculation}
%
%
\author{M. Karelin\\
 B. I. Stepanov Institute of Physics, National Academy of Sciences,
 \\
 Minsk, 220072 Belarus E-mail:~karelin@ifanbel.bas-net.by}
\date{}
\maketitle
\begin{abstract}
An entanglement measure for pure-state continuous-variable
bi-partite problem, the Schmidt number, is analytically calculated
for one simple model of atom-field scattering.
\end{abstract}


%
\section{Introduction}

Quantum entanglement --- non-classical correlations between parts
of compound quantum system is one the most important topics of
contemporary quantum theory, including quantum optics and quantum
information \cite{Bouwmeester_etal2000}.

One of the methods for analysis of bi-partite pure entangled
states is the Schmidt decomposition
\cite{Grobe_etal1994,EkertKnight1995,Eberly2005} ---
representation of the given state as a sum of product terms, where
the basic states are eigenvectors of marginal density matrices.
For continuous-variable (CV) problems, the Schmidt decomposition
gives an effectively finite-dimensional Hilbert space and
entanglement is characterized by the Schmidt number --- reciprocal
of the sum of marginal density operator eigenvalues.

In a recent series of publications, see further references in
\cite{Eberly2005,Torres_etal2005}, concerning the Schmidt
decomposition-based analysis of various problems, the entanglement
parameters were calculated mainly from numerical decomposition. In
fact, the only exception is the case of double-Gaussian wave
function with analytically known decomposition, see for example
\cite{Law2004,LawEberly2004,Wang_etal2006,LeeLaw2006} or Gaussian
mixed states \cite{Adesso_etal2004a}.

The aim of the present paper is to discuss another way of the
Schmidt number calculation --- from marginal purity. This method
is rarely mentioned in the published works, however in many cases it
enables to obtain the Schmidt number (in exact or approximate form)
analytically, and, for certain tasks, eliminate the need to calculate
the decomposition itself.

\section{CV pure-state entanglement and the Schmidt
decomposition}

The Schmidt decomposition, known also by different other names in
various scientific fields \cite{Holmes_etal1996}, in quantum
theory is an analysis of non-separable wavefunctions via
representation
 \begin{equation}
 \label{eq-1}
 |\varphi^{\mathrm(AB)}\rangle = \sum\nolimits_i \sqrt \lambda_i \;
     |\varphi_i^{\mathrm(A)}\rangle \; |\varphi_i^{\mathrm(B)}\rangle,
 \end{equation}
where the state vectors $|\varphi_i^{\mathrm(A)}\rangle$ and
$|\varphi_i^{\mathrm(B)}\rangle$ forms orthonormal systems
belonging to different parts of the composite quantum system and
all $\lambda_i \geqslant 0$. These state vectors are eigenvectors
of marginal density matrices of sub-systems, and coefficients
$\lambda_i$ are corresponding eigenvalues. According to the
decomposition (\ref{eq-1}), different Schmidt modes appears in
parts of a system in pairs, reflecting correlations between
sub-systems.

For CV systems, there is in principle infinite countable set of
eigenvalues $\lambda_i$. However, for all quantum
systems, sum of all eigenvalues 
is essentially finite, therefore only the limited number of
eigenvalues has significant value, and the whole entangled system
becomes effectively finite-di\-men\-sional \cite{LawEberly2004}.

The number of significant terms in representation (\ref{eq-1}) can
be characterized by the value
 \begin{equation}
 \label{eq-2}
 K = 1/\sum\nolimits_i \lambda_i^2,
 \end{equation}
effective number of the Schmidt modes (or the Schmidt number). On
the other hand, the Schmidt number is expressed via trace of
squared marginal density matrix ($\Tr_A$ and $\Tr_B$ are partial
traces over respective sub-system)
 \begin{equation}
 \label{eq-3}
 K = 1/\Tr\left(\rho_A^2\right) = 1/\Tr\left(\rho_B^2\right),
 \end{equation}
 \begin{equation}
 \label{eq-3a}
 \begin{split}
 \rho_A & = \Tr_B\left(|\varphi^{\mathrm(AB)}\rangle
     \langle\varphi^{\mathrm(AB)}|\right), \\
 \rho_B & = \Tr_A\left(|\varphi^{\mathrm(AB)}\rangle
     \langle\varphi^{\mathrm(AB)}|\right).
 \end{split}
 \end{equation}

Eigenvector decomposition is optimal from several points of view,
see detail in \cite{Fu1968}, but it can be calculated analytically
only in several special cases. Despite the mathematical methods to
treat this problem effectively \cite{LamataLeon2005}, it is
desirable to deal at least with entanglement degree analytically.


Schmidt number is closely related to ``generalized entropies'' (or
generalized purities), proposed by R\'enyi, Tsallis
\cite{Tsallis1988} and others, see further references in
\cite{Dodonov2002,Karelin2005,Adesso_etal2004a}:
 \begin{equation}
 \label{eq-4}
 S_p = \frac{1 - \Tr\rho^p}{p - 1},\quad
 S_p^R = \frac{\ln\,\Tr\rho^p}{p - 1},\quad p > 1
 \end{equation}
($\rho$ is a marginal density matrix). Here the limit $p \to 1$
leads to usual Shannon-von Neumann entropy, standard parameter for
entanglement charactetization \cite{Bennett_etal1996} and the case
$p = 2$ (purity, or ''linear entropy'') gives the Schmidt number
 \[
 S_2 \equiv S_L = 1 - 1/K.
 \]
The Schmidt number has a special meaning. First of all, it can be
measured in a single-photon counting experiment, see
\cite{LawEberly2004,Adesso_etal2004a,Adesso_etal2004b} and the
references therein. From theoretical point of view, for the whole
family of parameters (\ref{eq-4}), the Schmidt number is most easy
to calculate, just using the formula (\ref{eq-3}), without
knowledge of the decomposition eigenvalues.

\section{Atom-photon scattering: a model}

An entangled atom-field wavefunction in dimensionless momentum
representation in simplest case depends on only one parameter
\cite{Chan_etal2002,Eberly_etal2003a,Eberly_etal2003b,Chan_etal2003}
 \begin{equation}
 \label{eq-5}
 C(k,q) = \frac{N \, \exp(- \delta q^{2}/\eta^2)}{\delta k + \delta q + i},
 \quad
 \eta = \frac{\hbar \omega_0 \,\sigma}{M c \gamma}
 \end{equation}
where  $\eta$ --- the control parameter, ratio of thermal
(motional) line broadening $\hbar \omega_0\, \sigma/(M c)$ to a
natural linewidth $\gamma$, $N$ is a normalization constant.

For this model, in the Raman scattering regime \cite{Eberly_etal2003b},
control parameter value can be as large as $\eta _R \approx 4500$,
opening a way of experimental realization of high-entanglement regime.

In the papers
\cite{Chan_etal2002,Eberly_etal2003a,Eberly_etal2003b,Chan_etal2003},
numerical treatment is applied for the Schmidt decomposition,
leading to following expression for $K(\eta)$ at large $\eta$
(result of numerical fitting)
 \begin{equation}
 K = 1 + 0.28\,\left(\eta - 1 \right).
 \label{eq-6}
 \end{equation}

On the other hand, a straightforward analytical integration
according to (\ref{eq-3a}) gives marginal (atom) density function
 \begin{equation}
 \label{eq-7}
 \rho(q_1, q_2) =
 \frac{\pi \, N^2 \, \exp( - \delta q_1^2/\eta^2 - \delta q_2^2/\eta^2)}
   {(\delta q_1 - \delta q_2)/2i + 1},
 \end{equation}
together with a normalization constant $N^2 =
\sqrt{2}\big/(\pi^{3/2}\, \eta)$.

Another step of integration according to (\ref{eq-3}) leads to
analytical formula for the Schmidt number
 \begin{equation}
 \label{eq-8}
 K = \frac{\eta}{2 \sqrt{\pi}}
 \frac{\exp(-4/\eta^2)}{1 - \erf(2/\eta)},
 \end{equation}
which has simple asymptotic form for $\eta \gg 1$
 \begin{equation}
 \label{eq-9}
 K = \frac{2}{\pi} + \frac{\eta}{2\sqrt{\pi}}.
 \end{equation}

\begin{figure}
 \label{fig1}
 \begin{center}
 \scalebox{0.7}{\includegraphics{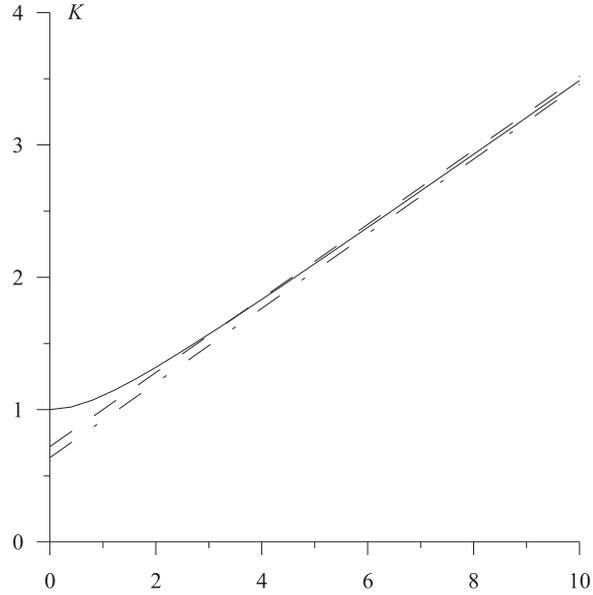}}
 \end{center}
 \caption{Different methods of the Schmidt number calculation.
Solid line --- exact dependence (\ref{eq-8}), dash-dot ---
asymptotic (\ref{eq-9}), dashed --- Eberly's approximation
(\ref{eq-6}).}
\end{figure}

Resulting dependencies $K(\eta)$ for these three expressions are
presented in Figure~\ref{fig1}. It is seen, that all the graphs
are quite close to each other.

%

\section{Conclusion}

The main task of the present paper --- to provide a simplest
example of an already known, but not widely used method of the
Schmidt number calculation. The Schmidt decomposition and the
Schmidt number prove to be quite efficient method for
characterization of pure state bi-partite entanglement, however,
its analytical calculation is possible just in a very limited
number of cases. Among the used simple example, analytical (or
semi-analytical) expressions can be found for another cases, for
example, for more generalized two-parametric atom-field
entanglement model \cite{GouGou2006}.

\section*{Acknowledgement}

The work has been partially supported by Belarusian Fund for
Fundamental Research, project F05K-056.


%
\end{document}